\documentclass{article}
\usepackage{spconf,amsmath,graphicx}

\usepackage{enumitem}
\setlist{nosep, leftmargin=14pt}
\usepackage{xcolor}     

\usepackage{mwe} 
\usepackage{booktabs} 

\title{From Global Radiomics to Parametric Maps: A Unified Workflow fusing Radiomics and Deep learning for PDAC Detection}
\name
 {Zengtian Deng$^{\star \dagger}$, Yimeng He$^{\star}$, Yu Shi$^{\star \dagger}$, Lixia Wang$^{\star}$, Touseef Ahmad Qureshi$^{\star}$, Xiuzhen Huang$^{\star}$, Debiao Li$^{\star \dagger}$}
 
\address{$^{\star}$ Cedars-Sinai Medical Center \\
    Los Angeles, CA, USA \\
    $^{\dagger}$University of California, Los Angeles \\
    Los Angeles, CA, USA\\
{\small Equal contribution (co-first): Zengtian Deng; Yimeng He.} \\
{\small Corresponding author: Debiao Li.}
}

\begin{document}
%

\maketitle
{\copyright~This work has been submitted to the IEEE for possible publication. Copyright may be transferred without notice, after which this version may no longer be accessible.}
\begin{abstract}

Radiomics and deep learning both offer powerful tools for quantitative medical imaging, but most existing fusion approaches only leverage global radiomic features and overlook the complementary value of spatially resolved radiomic parametric maps. We propose a unified framework that first selects discriminative radiomic features and then injects them into a radiomics-enhanced nnUNet at both the global and voxel levels for pancreatic ductal adenocarcinoma (PDAC) detection. On the PANORAMA dataset, our method achieved AUC = 0.96 and AP = 0.84 in cross-validation. On an external in-house cohort, it achieved AUC = 0.95 and AP = 0.78, outperforming the baseline nnUNet; it also ranked second in the PANORAMA Grand Challenge. This demonstrates that handcrafted radiomics, when injected at both global and voxel levels, provide complementary signals to deep learning models for PDAC detection. Our code can be found at https://github.com/briandzt/dl-pdac-radiomics-global-n-paramaps.
\end{abstract}
\section{Introduction}
\label{sec:intro}

Radiomics provides a quantitative framework for medical image analysis using handcrafted intensity, texture, and shape descriptors with strong interpretability and reproducibility \cite{aerts2014decoding,lambin2017radiomics,1_gillies2016radiomics,parmar2015machine}. Although deep learning often achieves superior end-to-end performance, purely data-driven models can be less interpretable and vulnerable to scanner and site specific domain shift \cite{lambin2017radiomics,kilim2022physical}. Accordingly, recent work has explored radiomics--deep learning fusion to combine handcrafted priors with learned representations \cite{li2025comparison}. However, most methods rely on global radiomic vectors and underutilize spatially resolved radiomics parametric maps \cite{kim2021radiomics,jensen2023role}. Existing voxel-wise approaches either compute exhaustive maps for many descriptors (high computational cost) or compress maps via singular-value decomposition (SVD), which can reduce feature-level interpretability \cite{van2017computational,chen2023radiomics}.

In this work, we present a novel application of radiomics features in deep learning framework for pancreatic ductal adenocarcinoma (PDAC) detection that jointly leverages global and voxel-level radiomic features. Our contributions are:

\begin{itemize}
\item \textbf{Unified radiomics--DL workflow.}
We first identify discriminative radiomic features through global analysis, then inject the selected features into deep learning as both case-level vectors and voxel-wise parametric maps, bridging global biomarkers with spatial cues for PDAC detection.

\item \textbf{Radiomics-enhanced nnUNet.}
We augment nnUNet by concatenating selected parametric maps with CT input and fusing global radiomics at the bottleneck via radiomics-aware cross-attention to improve lesion sensitivity and robustness.

\item \textbf{CUDA-accelerated parametric-map extraction.}
We implement a GPU-based voxel-level radiomics extractor using PyTorchRadiomics~\cite{liang2025localized}, substantially reducing per-feature extraction time and enabling large-scale parametric-map generation.
\end{itemize}


\section{Related Work}
\label{sec:related}

\subsection{Radiomics-Enhanced Deep Learning}
\label{sec:radiomics_in_deep_learning}



Radiomics provides a quantitative bridge between medical imaging and phenotype characterization \cite{aerts2014decoding,lambin2017radiomics,1_gillies2016radiomics,parmar2015machine}. Seminal work by Aerts et al. (2014) and Lambin et al. (2017) established that handcrafted radiomic descriptors capture tumor heterogeneity and predict clinically relevant outcomes \cite{aerts2014decoding,lambin2017radiomics}. 

Recent studies have explored integrating information, including radiomics, into deep learning (DL). Li et al. (2024) categorized information fusion in classification into input, output, and intermediate paradigms, and observed that most works follow single-layer intermediate fusion, which fused information after deep learning feature extraction but before final decision layer\cite{li2024review}. Within the scope of radiomics, both intermediate fusion and output fusion improved diagnosis and staging performance across modalities and cohorts \cite{lu2024deep,cai2024deep,liu2025pandx}. However, existing methods still treat radiomics as a global descriptor and typically fuse it at output, with limited mechanisms to (i) link global biomarkers to spatially resolved cues and (ii) pinpoint radiomic features critical for the problem.

Our approach addresses these limitations by using global radiomics as a \emph{selection and supervision signal}: discriminative global radiomic biomarkers are first identified, and only this compact subset is then integrated into the network through both global radiomics embedding and regional voxel feature maps as additional input channel. This dual-level, selection-linked design enables radiomics to function as plug-in ``feature adapters'' for standard segmentation backbones, rather than a one-off single-level fusion of global descriptors.

\subsection{Application of Radiomics Parametric Maps}
\label{sec:deep_learning_in_radiomics}


Radiomics parametric maps (voxel-wise radiomics) have emerged as a way to spatially resolve handcrafted descriptors and interpret them as images \cite{kim2021radiomics,jensen2023role,2_Lin2024_VoxelTextureSimilarityNetworks}. Kim et al. (2021) introduced accessible tooling for generating feature-specific maps, enabling radiomic features to be visualized and quantified locally \cite{kim2021radiomics}. Jensen et al. (2023) showed that map-first computation can improve robustness to varying region-of-interest (ROI) definitions, and Lin et al. (2024) extended voxel-level radiomics to 3D texture similarity networks to capture subject-level phenotypes from spatial feature distributions \cite{jensen2023role,2_Lin2024_VoxelTextureSimilarityNetworks}. 

Despite these advances, existing parametric-map studies are predominantly used for visualization or post hoc aggregation, and rarely serve as structured inputs to deep neural networks. Moreover, when maps are used, prior work typically lacks an explicit link to conventional global radiomics, leaving unclear which feature maps are most relevant and linked to learned global representations.

In contrast, we explicitly bridge global and voxel-wise radiomics: global radiomics as guided clue to pinpoint a compact subset of discriminative features, and only these selected features are materialized as parametric maps for prior information. Together with global radiomics embeddings fused via latent cross-attention, this yields a computationally feasible and interpretable mechanism for incorporating radiomics into segmentation backbones, aligning global biomarkers with their spatial realizations for lesion-aware detection.

\subsection{Deep Learning Based Pancreatic Cancer Detection}
\label{sec:deep_learning_based_pancreatic_cancer_detection}

Deep learning has rapidly advanced computer-aided pancreatic cancer detection and segmentation. Large-scale efforts such as the PANDA study (Cao et al., 2023) and nationwide validation by Chen et al. (2023) established high-performance CT-based models for PDAC screening, though most remain purely deep learning driven\cite{cao2023large,chen2023pancreatic}. nnUNet-based pipelines have become the de facto standard for pancreas and lesion segmentation, with subsequent work improving data scale and evaluation consistency \cite{yang2024nnu}. However, these image-only systems still face challenges in lesion sensitivity and cross-scanner variability. 

Our study addresses these gaps by injecting global radiomics analysis through radiomics-aware attention and voxel-level parametric mapping as additional channels to the UNet backbone, enhancing PDAC lesion detection accuracy and robustness.

\begin{figure}[htb]
  \begin{minipage}[b]{\linewidth}
    \centering
    \centerline{\includegraphics[width=8.0cm]{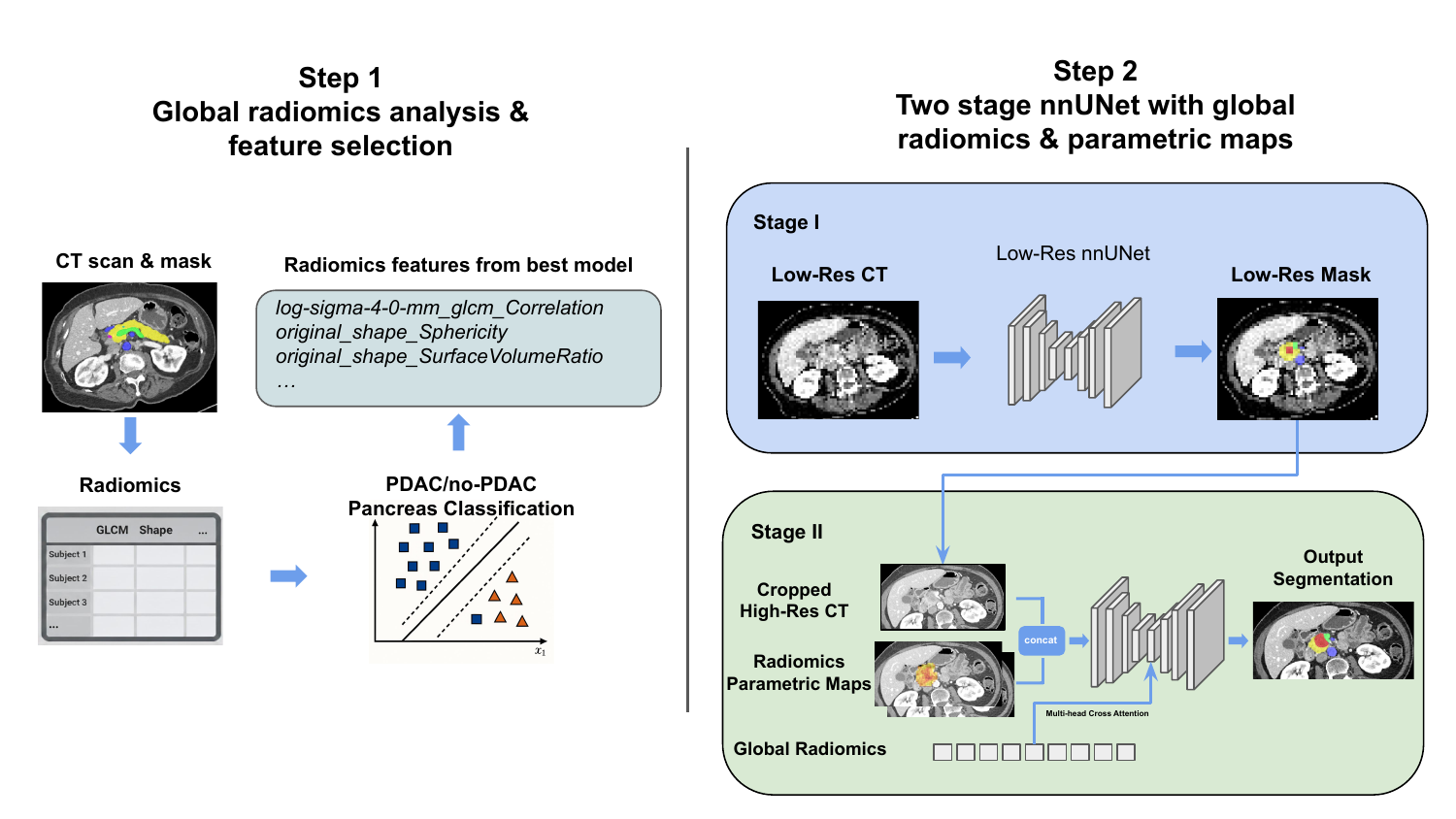}}
  \end{minipage}
  \caption{(Step 1) Whole-pancreas radiomics are extracted for PDAC classification, and the most informative features are selected. (Step 2) A coarse-to-fine two-stage network: Stage I predicts a rough mask to define the ROI and compute global/local radiomics; Stage II concatenates radiomics parametric maps with CT and fuses global radiomics via latent multi-head cross-attention to produce the final PDAC segmentation/detection.}
  \label{fig:res}
  \end{figure}

\section{Method}
\label{sec:method}

We proposed a radiomics-aware PDAC detection pipeline that combined offline feature discovery with a two-stage detector. First, we performed radiomics analysis over the whole pancreas to select discriminative features with case-level PDAC detection(§3.2). Then, we train a two-stage nnUNet: Stage 1 localizes the pancreas on low-resolution CT, and Stage 2 operates on the cropped high-resolution ROI for fine detection. In addition, Stage 2 network fused the selected radiomics in two forms — parametric maps by channel-wise concatenation at the input, and the global radiomics vector by cross-attention at the bottleneck (§3.3). This yields a unified model that exploits both voxel-level radiomics cues and case-level global descriptors for PDAC detection.

\subsection{Data Preparation}
\label{sec:data_preparation} 

We used the training dataset from the PANORAMA Challenge, which contains 2,238 venous-phase contrast-enhanced CT scans (1,562 non-PDAC and 676 PDAC). In addition, we curated an internal dataset from Cedars-Sinai Medical Center for external validation, consisting of 218 venous-phase contrast-enhanced CT scans (113 non-PDAC and 105 PDAC). Following Liu et al. (2025), we performed 5-fold stratified cross-validation based on lesion size (in voxels), using a bin size of 500 voxels to balance PDAC cases across folds \cite{liu2025pandx}.

\subsection{Global Radiomics Analysis}
Global Radiomics Analysis
We first performed a global radiomics analysis to identify discriminative case-level features that separated PDAC from non-PDAC scans. Using whole pancreas region including duct as the mask, We extracted 1{,}486 radiomics features (original image and variants including LoG, wavelet, etc.) with PyRadiomics~\cite{van2017computational}. The features spanned first-order category and common texture families (e.g., GLCM, GLRLM, etc.) and shape. All features were z-score standardized using statistics computed on the training folds only.  

To eliminate redundant features, we adopted a two-stage pipeline: (i) a univariate filter that retained features with significant two-sided Pearson correlation between groups under FDR control, followed by (ii) recursive feature elimination with an SVM on the retained set to filter features until only 10 features remained (e.g., GLCM Correlation, GLCM IMC1, NGTDM Strength, Shape Sphericity, Surface-Volume Ratio). Selection performance was estimated with 10-fold cross-validation on the training data, and the feature set from the best CV fold was chosen as the selected global radiomics features. From the selected 10 features, we excluded shape-based descriptors and generated voxel-wise parametric maps for the remaining 8 features with PyTorchRadiomics using a sliding-window kernel of 5 voxels for downstream use.

\subsection{Two-stage nnUNet for PDAC detection}

We adopted a two-stage, coarse-to-fine PDAC detection pipeline. At a high level, the first network localized the pancreas on a downsampled CT, and the second network focused on the cropped high-resolution region to detect the lesion and related structures with the help of the selected global radiomics features and their corresponding parametric maps.

\noindent \textbf{Stage 1 (Coarse Localization)} We trained an nnUNet on low-resolution volumes (4.5~$\times$~4.5~$\times$~9.0~mm) to obtain a pancreas prediction. At inference, this prediction was used to define a fixed ROI of 100~$\times$~50~$\times$~15~mm on the original CT volume.

\noindent \textbf{Stage 2 (Fine Segmentation/Detection)} We then trained a second nnUNet on the cropped full-resolution ROI. To make the detector more anatomically aware, we trained the model to jointly predict multiple structures — pancreas, abdominal aorta, portal vein, pancreatic duct, and common bile duct — rather than PDAC alone. This multi-structure supervision encouraged the model to learn the surrounding anatomy and improved robustness and generalization. In this stage, we also injected radiomics in two ways: (i) the voxel-wise radiomics parametric maps were fused by channel-wise concatenation with the CT at the input, and (ii) the case-level global radiomics vector was fused at the bottleneck via multi-head cross-attention, using the global radiomics vector as the query and the latent nnUNet features as keys and values. This combination allowed the network to use both local radiomics cues and global-level descriptors to improve detection performance. Both stages were trained with a combination of Dice loss and normalized cross-entropy.

For deployment realism, both the global radiomics and the parametric maps used by Stage~2 were computed from the Stage~1 pancreas prediction rather than ground-truth masks, during both training and inference.

To make voxel-wise radiomics practical for all cases, we generated the parametric maps with PyTorchRadiomics, which provided GPU-accelerated sliding-window extraction; the runtime comparison with vanilla PyRadiomics is reported in §4.3~\cite{liang2025localized,van2017computational}.

\section{Experimental Results}
\label{sec:results}

We conducted three sets of experiments. First, we evaluated the global radiomics module from §3.2 to confirm that the selected case-level features are individually discriminative for PDAC vs. non-PDAC (§4.1). Second, we assessed the full radiomics-aware two-stage nnUNet on both the PANORAMA dataset and our internal Cedars-Sinai cohort, and we performed ablations to isolate the contributions of global features and voxel-wise parametric maps (§4.2). Finally, we measured the runtime of the proposed PyTorchRadiomics implementation to show that voxel-wise radiomics extraction can be made practical for large-scale studies (§4.3).

\subsection{Global Radiomics-Based PDAC Classification}
\label{sec:PDAC_Detection}

Using the global radiomics features selected described in  §3.2, we trained an SVM to classify PDAC vs. non-PDAC on the PANORAMA training set. Evaluated with 10-fold cross-validation, the radiomics-only model achieved AUROC = 0.824 ± 0.035 and AP = 0.667 ± 0.046, indicating that the selected feature set carries meaningful discriminative signal. We therefore fixed this feature set and used it for the Stage-2 nnUNet.

\begin{table}[htbp]
  \centering
  \caption{5-fold cross validated PDAC detection on PANORAMA dataset}
  \label{tab:pancreatic_cancer_detection_on_panorama_dataset}
  \begin{tabular}{@{}lccc@{}}
  \toprule
  \textbf{Methods} & \textbf{AUROC} & \textbf{AP} & \textbf{(AUROC+AP)/2} \\
  \midrule
  nnUNet(baseline)      & $0.959$ & $0.810$ & $0.885$ \\
  + Global Features     & $0.957$ & $0.813$ & $0.885$ \\
  + Parametric Maps     & \textbf{$0.960$} & $0.826$ & $0.893$ \\
  + Both                & $0.958$ & $\textbf{0.836}$ & $\textbf{0.897}$ \\
  \bottomrule
  \end{tabular}
  \vspace{0.2em}
  \end{table}
  
\begin{table}[htbp]
  \centering
  \caption{PDAC detection on external test set}
  \label{tab:pancreatic_cancer_detection_on_inhouse_external_dataset}
  \begin{tabular}{@{}lccc@{}}
  \toprule
  \textbf{Methods} & \textbf{AUROC} & \textbf{AP} & \textbf{(AUROC+AP)/2} \\
  \midrule
  nnUNet(baseline)      & $\textbf{0.954}$ & $0.662$ & $0.808$ \\
  + Global Features     & $0.948$ & $0.715$ & $0.832$ \\
  + Parametric Maps     & $0.947$ & $0.743$ & $0.845$ \\
  + Both                & $0.951$ & $\textbf{0.777}^*$ & $\textbf{0.864}^*$ \\
  \bottomrule
  \end{tabular}
  \vspace{0.2em}
  \begin{minipage}{0.95\linewidth}
    \footnotesize
    \raggedright
    \textbf{*} p-value $<$ 0.01 when compared to baseline nnUNet.\\
  \end{minipage}
\end{table}

\begin{figure}[htb]
  \begin{minipage}[b]{.48\linewidth}
    \centering
    \centerline{\includegraphics[width=4.0cm]{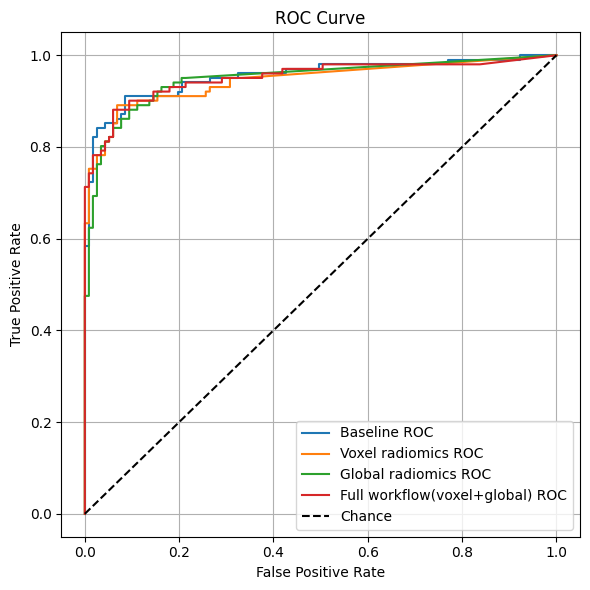}}
    \centerline{(a) Case-wise ROC curve}\medskip
  \end{minipage}
  \hfill
  \begin{minipage}[b]{0.48\linewidth}
    \centering
    \centerline{\includegraphics[width=4.0cm]{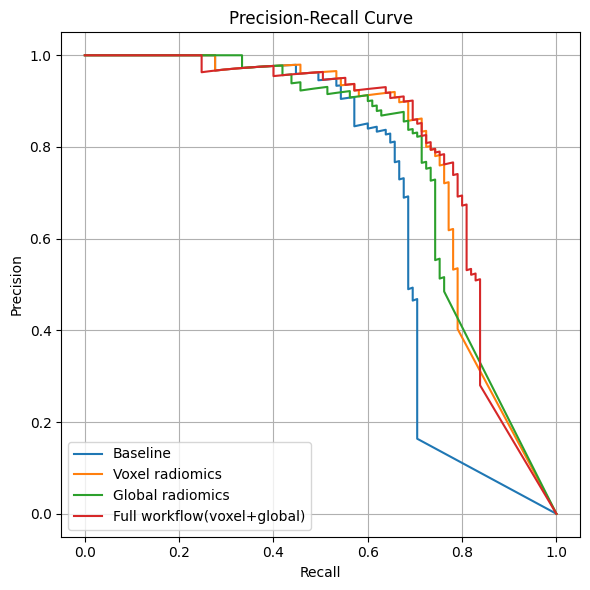}}
    \centerline{(b) Lesion-wise PR curve}\medskip
  \end{minipage}
  \caption{ROC and PR curves for in-house external dataset PDAC detection.Our model with combined global and local radiomics features achieves better performance (0.662 -> 0.777, p<0.01) than the baseline nnUNet model.}
  \label{fig:res}
  \end{figure}


\subsection{PDAC Detection}
\label{sec:pancreatic_cancer_detection}

We evaluated the proposed workflow on both the PANORAMA training dataset (5-fold cross-validation) and our external cohort. We used picai-eval to ensure consistent lesion-level Average Precision(AP) and case-level AUROC metrics extraction across datasets \cite{saha2024artificial}. To quantify the contribution of each radiomics component, we also performed ablation experiments in which we added only the global radiomics vector or only the voxel-wise parametric maps to the second-stage PDAC detector during training. Finally, we report our PANORAMA test-set result on the challenge leaderboard to compare with other participating teams.

As shown in Table 1 (PANORAMA) and Table 2 (External), combining global radiomics and parametric maps yielded best performance on both the cross-validated training data and the external test set. On the in-house cohort, the combined model also achieved a statistically significant improvement over the baseline (p = 0.002). In addition, based on both Table 1 and Table 2, the model with parametric maps only showed better performance than the model with only global radiomics, which was likely due to the noisy global information due to imperfect segmentation from Stage 1 nnUNet. For the PANORAMA test set evaluation, our preliminary model was trained with only three radiomics parametric maps without global radiomics fusion, but we still achieved 2nd place of overall ranking.

\subsection{Runtime of Voxel-Wise Radiomics}
\label{sec:speed_comparison}

To assess the practicality of voxel-wise radiomics at scale, we measured the extraction time for the 8 selected parametric maps on 50 cases using (i) vanilla PyRadiomics and (ii) our GPU-based PyTorchRadiomics implementation \cite{van2017computational,liang2025localized}. Based on the experiment, PyTorchRadiomics reduced the mean extraction time from 53.42s to 16.28s per feature, a substantial speed-up with $p \ll 0.01$, making per-case map generation feasible for large cohort.

\section{Conclusion}
\label{sec:conclusion}

In this work, we proposed a unified framework that first performed global radiomics analysis to identify discriminative features, then a two-stage nnUNet which integrated both selected global radiomics features and their respective parametric maps. We applied our framework to Pancreatic Adenocarcinoma (PDAC) detection task and achieved better performance than baseline nnUNet and ranks second in the PANORAMA Grand Challenge \cite{alves2024panorama}. Our work showed that conventional radiomics and parametric maps together can serve as complementary prior to improve robustness in downstream deep learning analysis, and by including gpu-accelerated radiomics parametric map extraction, our proposed workflow can be more scalable and efficient for real-world applications.

\section{ACKNOWLEDGMENTS}
\label{sec:acknowledgements}

This work was supported by the National Institutes of Health (NIH) under grant R01 CA260955.

\section{Compliance with ethical standards}
\label{sec:ethics}
The internal dataset of this study was retrospectively collected and de-identified at Cedars-Sinai Medical Center under an Institutional Review Board (IRB)–approved protocol and analyzed under a waiver of informed consent; The public dataset was obtained from the PANORAMA Challenge, which provides fully de-identified imaging data under an open-access license \cite{alves2024panorama}.

\newpage
\bibliographystyle{IEEEbib}
\bibliography{strings,refs}

\end{document}